\documentclass[%
reprint,
superscriptaddress,
 amsmath,amssymb,
 aps,
prl,
]{revtex4-1}

\usepackage{graphicx}% Include figure files
\usepackage{dcolumn}% Align table columns on decimal point
\usepackage{bm}% bold math
\usepackage{color}
\usepackage[colorlinks=true,linkcolor=blue,citecolor=blue,urlcolor=blue,pdfborder={0 0 0}]{hyperref} % add hypertext capabilities
\begin{document}

%\preprint{APS/123-QED}

\title{Superfluid flow above the critical velocity}
% Force line breaks with \\
%\thanks{A footnote to the article title}%

\author{A. Paris-Mandoki}
\affiliation{Midlands Ultracold Atom Research Centre, 
School of Physics \& Astronomy,
University of Nottingham, Nottingham NG7 2RD, United Kingdom}

\author{J. Shearring}
\affiliation{Midlands Ultracold Atom Research Centre, 
School of Physics \& Astronomy,
University of Nottingham, Nottingham NG7 2RD, United Kingdom}

\author{F. Mancarella}
\affiliation{Nordic Institute for Theoretical Physics (NORDITA), SE-106 91 Stockholm, Sweden}
\affiliation{Department of Theoretical Physics, KTH Royal Institute of Technology, SE-106 91 Stockholm, Sweden}

%\author{S. Fagnocchi}
%\affiliation{Midlands Ultracold Atom Research Centre, 
%School of Physics \& Astronomy,
%University of Nottingham, Nottingham NG7 2RD, United Kingdom}

\author{T.M. Fromhold}
\affiliation{Midlands Ultracold Atom Research Centre, 
School of Physics \& Astronomy,
University of Nottingham, Nottingham NG7 2RD, United Kingdom}

\author{A. Trombettoni}
\affiliation{CNR-IOM DEMOCRITOS Simulation Center, 
Via Bonomea 265 I-34136 Trieste, Italy}
\affiliation{SISSA and INFN, Sezione di Trieste, 
Via Bonomea 265 I-34136 Trieste, Italy}

\author{P. Kr\"uger}
\email{Peter.Kruger@nottingham.ac.uk}
\affiliation{Midlands Ultracold Atom Research Centre, 
School of Physics \& Astronomy,
University of Nottingham, Nottingham NG7 2RD, United Kingdom}

\date{\today}% It is always \today, today,

\begin{abstract}
Superfluidity and superconductivity have been studied widely since the last century in many different contexts ranging from nuclear matter to atomic quantum gases. 
The rigidity of these systems with respect to external 
perturbations results in frictionless motion for 
superfluids and resistance-free electric 
current in superconductors. 
This peculiar behaviour is lost when external perturbations overcome a critical threshold, i.e.\ above
a critical magnetic field or a critical current for superconductors. 
In superfluids, such as liquid helium  
or ultracold gases, 
the corresponding quantities are critical rotation rate and critical velocity, respectively. Enhancing the critical values is of great fundamental and practical value. Here we demonstrate that superfluidity can be achieved for flow above the critical velocity through quantum interference induced resonances. This has far reaching consequences for the fundamental understanding of superfluidity and superconductivity and opens up new application possibilities in quantum metrology, e.g.\ in rotation sensing.
\end{abstract}

% PACS, the Physics and Astronomy
                             % Classification Scheme.
%\keywords{Suggested keywords}%Use showkeys class option if keyword
                              %display desired
\maketitle

%\tableofcontents

% Superfluidity and Sperconductivity. Introduce breakdown. Importance of breakdown (abstract and practical)(mention superconductors).
 
%It is clear that having a mechanism to increase those 
%critical values (e.g., critical current in superconductors 
%and critical velocity in superfluids) or to have 
%superfluid/superconducting behaviour above them is of 
%fundamental and practical relevance. In this paper we study supercritical motion 
%in presence of defects with rectangular 
%shapes having a well definite length scale and we 
%report on the existence of a set of velocities 
%above the critical velocity for which the superfluid 
%motion is preserved and the production of excitations 
%is inhibited.

% Breakdown and critical velocity

\begin{figure}[t]
\includegraphics[width=\columnwidth]{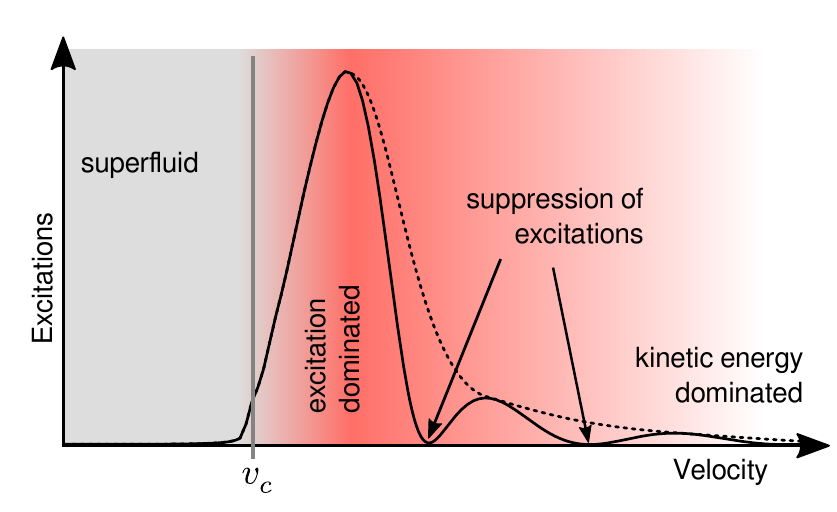}
\caption{Excitations in a superfluid as a function of velocity of a defect dragged through the fluid at a given time. 
For velocities below a critical velocity ($v<v_c$), the production of excitations is suppressed and superflow persists: 
at longer times the amount of excitations remains very small. 
At higher velocities ($v>v_c$) a sharp onset of  
excitations (growing with time) destroys the superfluid properties and only at very high velocities ($v\gg v_c$) 
the kinetic energy becomes so high that the defect hardly affects the flow. This standard picture (dashed line) has to be adjusted when resonant quantum interference reinstates superfluidity with fully suppressed excitation growth at a series of discrete supercritical velocities. The solid line shows an example for a rectangular defect shape
(for the same plot at different times see Fig.\ \ref{fig2}).}
\label{fig1}
\end{figure}

The breakdown of 
superfluidity and superconductivity above
the critical velocity and the critical current \cite{Tinkham96,Leggett06}, respectively, 
is caused by the production and growth of excitations.
In a superfluid, when its flow velocity, or equivalently the velocity of a 
defect dragged through the fluid, exceeds the critical velocity $v_c$ 
the creation of excitations becomes energetically 
favourable. This destroys the frictionless 
motion, as shown in experiments with 
superfluid helium \cite{Allum77}, and
ultracold bosons \cite{Raman99} and fermions 
\cite{Miller07}. 
The critical velocity is defined as 
the maximum velocity below which there is no (or more precisely a bounded) production  
of excitations.
%This standard scenario of the breakdown of superfluidity is depicted in Fig.\ \ref{fig1} where 
%the produced excitations (at a given time) are plotted as a function of the fluid velocity (solid line). 
While (essentially) no excitations are present for subcritical velocities, a fast onset of growing excitations occurs for supercritical velocities that gradually decreases for further increased velocity until the kinetic energy of the fluid becomes so high that it dominates all other energy scales and any defects become unimportant again (Fig.\ \ref{fig1}).

In general, the rate of excitation growth depends 
on the microscopic details of the superfluid and 
on how it is coupled to the environment. 
The details of this coupling determine the dissipation mechanism 
causing the creation of excitations and 
ultimately the critical velocity $v_c$. A simple way to estimate 
$v_c$ and give a qualitative explanation 
of the breakdown of superfluidity is provided by the Landau criterion \cite{Lifshitz80}.

% Landau criterion
% Landau give a qualitative and a rough quantitative in 1D. In general give upper bound. It is not necessary or sufficient. Perturbative.

The Landau criterion is a cornerstone of our understanding 
of the dynamical behaviour of superfluids, stating 
that a superfluid flow is sustained against external perturbations or defects up to a critical 
value of the velocity \cite{Lifshitz80}. 
Its elegance, power and usefulness 
rely both on simplicity and generality: 
there is no need to know the specific nature of the
perturbation or the characteristics 
of the defects, no need to know the microscopic details of the superfluid, and no need to compute the excitation spectrum of the moving system; only knowledge of the 
low-energy excitation spectrum $\epsilon(p)$ 
of the system at rest is required. 
Briefly, by applying a
Galileo transformation to the co-moving frame it can be shown that 
for $v<v_L$ (where $v_L=\min{\frac{\epsilon(p)}{p}}$ 
is the Landau critical velocity \cite{Lifshitz80}) 
the production of elementary excitations is energetically 
unfavoured. From the Landau criterion it follows that for short-range weakly interacting Bose gases, the critical velocity 
is equal to the sound velocity $c$, so that here ``supercritical'' means ``supersonic''. 
Superfluidity in a weakly interacting Bose gas flowing at a velocity greater than the
Landau critical velocity was studied in \cite{Baym12}.
Although the detailed analysis of different 
superfluid systems, including helium and ultracold gases, 
shows that the Landau criterion often quantitatively overestimates the actual critical velocity $v_c$, 
especially in the two- and three-dimensional case
(see also a recent paper on one dimension \cite{Finazzi14}),  the identification of a critical 
velocity above which the production of excitations destroys the superfluid motion is a criterion of paramount 
clarity and relevance.

The Landau criterion is based on a perturbative treatment of ``small'' defects
affecting the superfluid motion. The possibility 
to explore superfluid motion in a non-perturbative regime of parameters has attracted considerable interest, e.g.\ in non-perturbative and/or exact studies of the dynamical 
propagation of a superfluid in presence of ``non-small'' 
defects (of tunable shape and intensity) or periodic potentials \cite{Ianeselli06}. 
The point we address in this paper is the very surprising possibility of stable {\em supercritical}
propagation (i.e.\ {\em above} the critical velocity) at a non-perturbative level. 
In the following we show that it is possible to have a set of 
velocities {\em larger} than the critical velocity $v_c$ where superfluid flow exists and the production of excitations is completely suppressed. The key mechanism here is a recurring resonant phenomenon between the nonlinear wave propagation of the superfluid and the defect. The resulting scenario is plotted 
in Fig~\ref{fig1} as a solid line, in contrast to the very general notion of the absence of superfluidity for $v>v_c$ (dashed line).

% Setup

% What others did
Superfluid motion in presence of defects has been extensively investigated 
in ultracold atoms: superfluidity can be probed 
by stirring a laser beam \cite{Raman99,Onofrio00,Engels07,Neely10} and the 
critical velocity has been measured \cite{Raman99}. Experiments on superfluid 
motion have been performed 
also with moving optical lattices \cite{Mun07}, in toroidal geometries 
\cite{Ramanathan11,Moulder12}, with ultracold fermions 
near unitarity \cite{Miller07} and in two-dimensional Bose systems \cite{Desbuquois12}. 
%Maybe move to applications. It is out of place here.

To illustrate the existence of supercritical flow and the associated arising phenomena we choose defects of rectangular shape in the one-dimensional ($1d$) flow of a superfluid, whose behaviour is governed by the Gross-Pitaevskii equation.  We consider a homogeneous system with stationary flow at velocity $v$ in an initially flat potential, in which a rectangular defect is then introduced to study the dynamical response. 
From a theoretical perspective, a challenging point is to define the transmission and reflection coefficients since the usual 
definitions used for linear matter waves \cite{Cohen77,Gilmore04} do not apply anymore; in fact the superposition principle of an incoming and a reflected wave is no longer valid, and bound states 
in the defect can be present due to the interaction term \cite{Leboeuf01}. 
For the interacting gas one can quantify the transmitted part of an incident wavepacket 
\cite{Stiessberger00,Paul05,Paul07,Kamchatnov08,Leszczyszyn09,Watanabe09,Piazza11} 
or characterize the breakdown of superfluidity caused by the drag exerted by a matter wave on an obstacle 
\cite{Pavloff02}. In presence of a $\delta$-like potential moving at supersonic velocity the stationary wave patterns 
have been studied \cite{Horng09}.
When the barrier is rectangular, solutions of the time-independent Gross-Pitaevskii equation can be written in terms of Jacobian 
functions \cite{Carr05,Rapedius08}: for a rectangular well the current-phase relation 
for subsonic motion can be determined \cite{Baratoff70,Piazza10} and when neglecting the mean-field interaction outside the potential 
well, it is possible to analytically calculate the transport properties of the system in terms of incoming and outgoing waves and resonances and bound states are obtained in closed form \cite{Rapedius06}. 
Understanding matterwave propagation in the presence of tailored 
defect potentials is important for the study of analogue gravity models and acoustic Hawking radiation in Bose-Einstein condensates \cite{Barcelo05}.
%\cite{Carusotto08,Recati09,Finazzi11,Larre12,Steinhauer14,Boiron14}, relevant 

It is well known that in the non-interacting regime the transmission coefficient across a square potential, as found by solving the time-independent {\em linear} Schr{\"o}dinger equation, reaches exactly unity for specific values of the momentum of an incident plane wave 
(Ramsauer-Townsend resonances). 
In this work we quantify the creation of excitations, given the geometric parameters of the defect, 
in relation to the flow velocity $v$ and the interaction strength $g$ in the gas. We find that the resonant behaviour has a counterpart in the {\em nonlinear} regime and superfluid motion across a defect is possible in spite of the Landau criterion predicting growth of excitations.
The velocities at which this excitation-free supercritical flow occurs are shifted and continuously connect to the velocities of unit transmission in the linear case as the mean-field interaction parameter is varied. This can be viewed as generalised Ramsauer-Townsend resonances that continuously connect the non-interacting case to the attractively and repulsively interacting regimes. 

Given the explicit arguments of 
the Landau criterion it is important to clarify what is meant by supercritical superfluid flow: {\em i)} A supercritical solution of the dynamical equations exists for specific values of the superfluid's momentum and system parameters (i.e., interaction strength and geometric properties of the defect) and the flow of such a solution is stable under small perturbations.
{\em ii)} The perturbations of the propagating flow created by the defect is bounded in time  
(at least for experimental timescales) and no new excitations are produced when the barrier is completely ramped on. 
%{\em iii)} The flow should be stable under 
%small perturbations: 
%in other words, small perturbations should not destroy the boundedness of the 
%growth of excitations. 
{\em iii)} The superfluid flow is physically inducible and accessible. We have found that all three conditions are fulfilled for the supercritical flow discussed here.

We have verified that the production of excitations at the supercritical excitation-free points 
is bounded in time, that the flow is stable under small perturbations 
and for times larger than 
the typical experimental timescales of ultracold atom experiments 
with $1d$ Bose gases, and that the obtained findings do not crucially depend 
on the ramping time of the well/barrier.
%A 
%comparison with multiple-scale analysis for small nonlinearities \cite{Ishkhanyan09} 
%and a discussion on a possibile experimental implementation are presented as well. 

% Model GPE V(x,t)
The Gross-Pitaevskii equation we consider is 
\begin{equation}
i \hbar \frac{\partial \psi}{\partial t}=-\frac{\hbar^2}{2m} \frac{\partial^2 \psi}{\partial x^2} + V(x,t) \psi + g |\psi|^2 \psi 
\label{GPE}
\end{equation}
where $\psi(x,t)$ is the condensate wavefunction, $g$ is the one-dimensional nonlinear coefficient \cite{Olshanii98}, and the potential is chosen 
to be   
\begin{equation}
V(x,t)=\left\{ \begin{array}{ll} 
V_0 \tanh{\left( \frac{t^2}{\alpha^2}\right)} &        \mathrm{for} \, \, \, 0<x<d\\
0 & \mathrm{otherwise}
\end{array} \right. ,
\label{barrier}
\end{equation}
where $d$ is the width and $V_0$ the strength of the defect, which is ramped on at time $t=0$ with a speed parametrized by $\alpha$. At time $t=t_{\mathrm{barrier}}\equiv1.5\alpha$ 
the value of the defect is $0.98V_0$, so it is practically almost completely turned on. The initial state (when the barrier is absent) 
is a plane wave with momentum $k$ and velocity $v=\hbar k /m$: $\psi(x,t=0)=\psi_0 e^{ikx}$ and density $n\equiv|\psi_0|^2$. Since for $t=0$ the defect is absent, $\psi(x,t=0)$ is a solution with momentum $k$ of the time-independent Gross-Pitaevskii equation. The idea of ramping on the defect is to adiabatically lead the system towards possible superctical superfluid solutions for specific values of the momentum $k$.

We consider periodic boundary conditions on a domain with finite length $2L \gg d$ (with $x \in [-L,L]$), and we check that for the considered times the effects of raising the barrier do not propagate to the boundaries; thus, the results are independent of the specific choice 
of boundary conditions. For the numerical solution we used the projected fourth-order Runge-Kutta method in the interaction picture \cite{Norrie05}, 
which we confirmed to be very stable for all times within the considered range.  

\begin{figure}[t]
\includegraphics[width=\columnwidth]{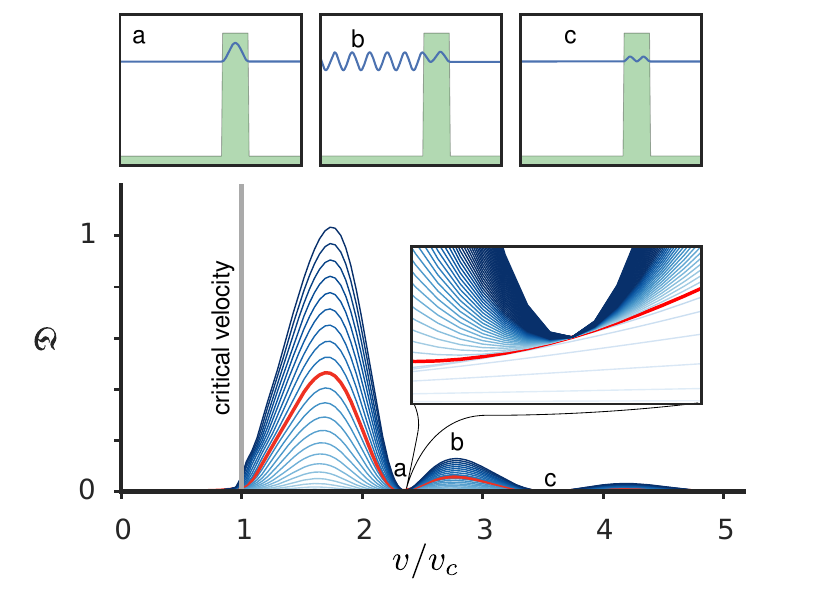}
\caption{Disturbance $\mathfrak{D}$ vs $v/v_c$ at various equidistant times (growing from the bottom of the figure), where $v_c$ is equal to the sound velocity $c$. Darker lines indicate longer times. The line highlighted in red corresponds to $t=t_{\mathrm{barrier}}$. At specific values of $v$, the disturbance does not grow with time after the barrier has finished rising. The insets illustrate this phenomenon in detail for the first minimum. Panels (a),(b),(c) show the density $n(x)=|\psi(x,t_0)|^2$ (in blue) and the potential $V(x, t_0)$ (in green) at time $t_0 =2.2 t_{\mathrm{barrier}}$ for three different initial values of $v$ corresponding to the minimum and maximum points for $\mathfrak{D}$ indicated as ``a'', ``b'', ``c'' in the main figure. The initial condition is a plane wave travelling with momentum $\hbar k=mv$ towards the right. The simulation parameters are $g = 15\times\left(\hbar^2/2md\right) $, $\alpha= 3\times \left(2md^2/\hbar\right) $ and $V_0 = 2 \times \left( \hbar^2 / 2md^2 \right)$.}
\label{fig2}
\end{figure}

\begin{figure}[t]
\includegraphics[width=\columnwidth]{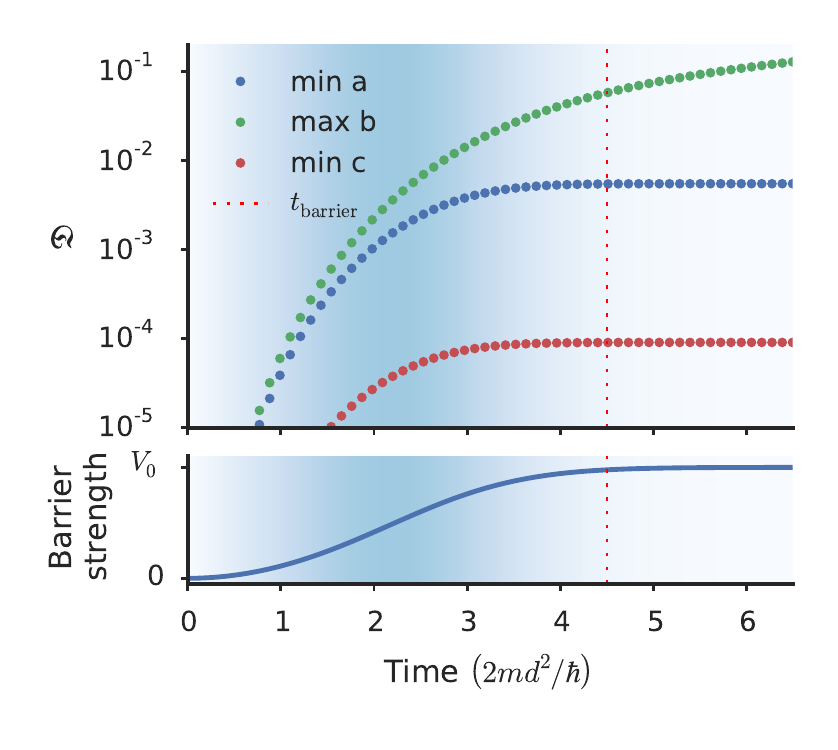}
\caption{Disturbance $\mathfrak{D}$ as a function of time for the initial velocities 
$v$ corresponding to the minima and maximum indicated in Fig.\ \ref{fig2} (top) - the bottom plot 
shows the corresponding time dependence of the barrier strength. The background shading indicates the rate of change of the barrier strength. For times $t>t_\mathrm{barrier}$, where the barrier strength has settled, $\mathfrak{D}$ has a linear dependence on time. The creation of excitations can be characterised by the 
slope of the disturbance $\dot{\mathfrak{D}}$ at large times and it is extremely small at the minima (see central right panel in Fig.\ \ref{fig4} for details).}
\label{fig3}
\end{figure}

%\begin{figure}[t]
%\includegraphics[width=\columnwidth]{fig4.pdf}
%\caption{Disturbance $\mathfrak{D}$ as a function of time for various values of $\alpha$; ranging from instantaneously raising the barrier (dark red) to slowly increasing with a time-constant $\alpha = 5\times\left(2md^2/\hbar\right)$ (dark blue). The left and right panels correspond to the initial conditions a and b indicated in Figure~\ref{fig2} respectively. More excitations may be created if the barrier increases faster, however, the asymptotic rate of excitation creation $\dot{\mathfrak{D}}(t\rightarrow\infty)$ does not depend on the speed witch which the barrier is raised. }
%\label{fig4}
%\end{figure}

%\begin{figure}[t]
%\includegraphics[width=\columnwidth]{fig5.pdf}
%\caption{ }
%\label{fig5}
%\end{figure}

\begin{figure}[t]
\includegraphics[width=\columnwidth]{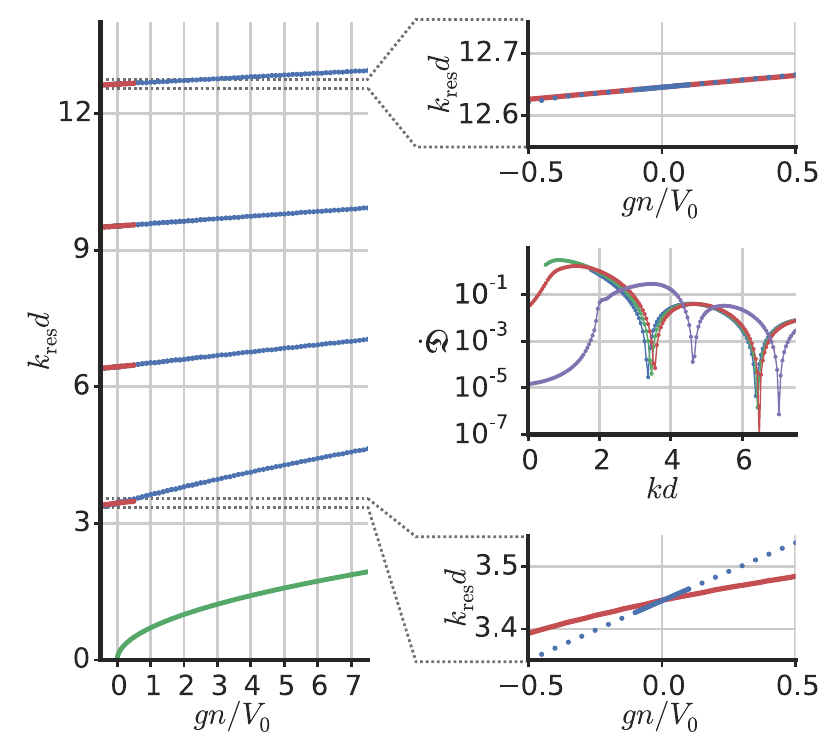}
\caption{Resonant momenta (with $ m v_\mathrm{res}=\hbar k_\mathrm{res}$). Centre-right: disturbance growth rate $\dot{\mathfrak{D}}$ 
as a function of $k d$ for four different values of interaction: $gn/V_0=$-0.5, 0, 0.5, 7.5 (blue, green, red, purple lines, respectively). Left: Resonant momenta $k_\mathrm{res}d$ 
as a function of $gn/V_0$ for the first four excitation-free points. The solid red
line is the perturbative prediction (Eq.\ \ref{multiple}), the blue points indicate numerically calculated values and the green line shows the sound velocity $c=\sqrt{gn/m}$. Bottom-right and top-right: detailed view of the first and fourth resonances, respectively. 
}
\label{fig4}
\end{figure}

% Results: write around the figures
To quantify the production and growth of excitations we found convenient to introduce the time-dependent quantity $\mathfrak{D}(t)$ 
(which we refer to as the disturbance) defined as 
\begin{equation}
\mathfrak{D}(t) = \int_{-L}^{0} dx \left( |\psi(x,t)|^2 - |\psi(x,t=0)|^2 \right)^2, 
\label{Ddot}
\end{equation}
where the integral is calculated over the region where the initial wave propagation is directed towards the defect. $|\psi(x,t=0)|^2$ is the density $n$. Of course, if the defect is absent $\mathfrak{D} = 0$.

Our results are summarized in Figs.\ \ref{fig2}-\ref{fig4}. Fig.\ \ref{fig2} shows that essentially no excitations 
are produced for velocities below the sound velocity $c=\sqrt{g n/m}$ (here $v_c\approx v_L=c$). As expected, excitations are produced 
for $v>v_c$, as time progresses $\mathfrak{D}(t)$ increases, and for large velocities $\mathfrak{D}(t)$ becomes smaller. However, 
there are velocities $v>v_c$, for which the production of excitations is inhibited. As shown in the inset 
of Fig.~\ref{fig2}, close to these points the production of excitations is bounded in time. Panels (a), (b), (c) of Fig.~\ref{fig2} show 
the density at a fixed time larger than $t_{\mathrm{barrier}}$. Away from the minima phonons are emitted (Fig.\ \ref{fig2}b) while at the minima a stationary breathing state forms inside the defect (Fig.\ \ref{fig2}a and Fig.\ \ref{fig2}c)\cite{Pavloff02}. 
We checked that these results neither depend on the particular choice of the measure of disturbance considering other 
quantifiers of the excitations, nor on the periodic boundary conditions or the choice of $L$. These 
findings also do not critically depend on the value $\alpha$. This is true for typical 
experimental values of $\alpha$ as well as for 
$\alpha \to 0$, where small fluctuations of $\mathfrak{D}(t)$ occur for specific choices of quantifiers of the disturbance. We finally observe that our 
results hold for a wide range of interactions, including large values of $g>0$, so that the validity spans the regime where
the healing length is smaller than the defect width to the regime where it is larger. For small 
negative values of $g$ the resonances are still present, but when $-g$ becomes large then collapse and 
instabilities are observed as expected.

As shown in Fig.\ \ref{fig3}, there is a clear difference 
between the growth of $\mathfrak{D}(t)$ at the resonant and non-resonant velocities. At resonant velocities, excitations are produced exclusively during the ramping of the defect and the disturbance is afterwards constant. For non-resonant velocities the disturbance grows linearly with time, which leads us to quantify the growth of excitations by computing 
the time derivative of $\mathfrak{D}(t)$ for $t>t_{\mathrm{barrier}}$ (and checking that the obtained value 
does not depend on the computation interval). The analysis shows that $\dot{\mathfrak{D}}$ is extremely small at the resonant velocities (being suppressed by at least 4 orders of magnitude with respect to non-resonant velocities). $\dot{\mathfrak{D}}$ is also very small for $v<v_c$ in agreement with the Landau 
criterion. In the inset of Fig.\ \ref{fig4} we plot the values 
$\dot{\mathfrak{D}}$ 
as a function of the velocity for four interaction strengths. By performing the same 
analysis we can reconstruct the behaviour of the resonant velocities $v_{\mathrm{res}}$ as a function 
of $g$: the results are plotted in Fig.\ \ref{fig4} for a barrier (positive $V_0$). With respect to the 
non-interacting limit $g=0$ the shift is positive (negative) for repulsive (attractive) interactions $g>0$ ($g<0$). Similar results hold for a potential well ($V_0<0$), then with a negative (positive) shift for repulsive (attractive) interactions. Our results are plotted 
in Fig.\ \ref{fig4} together with a multiple-scale analytical derivation valid for small $g$, predicting values of $k_\mathrm{barrier}$ (defined below) corresponding to a total transmission across the barrier \cite{Ishkhanyan09}.  These momenta are given by $k_{\mathrm{barrier}}d=n\pi+\delta$, where 
$k_{\mathrm{barrier}}=\sqrt{k^2-\frac{2mV_0}{\hbar^2}}$ (with velocity $v=\hbar k/m$) and 
\begin{equation}
\delta=\frac{3a(kd+n\pi)}{8}
\label{multiple}
\end{equation}
with $a=2mgn/v^2$. Our result differs by the factor $(-1)^n$ from Eq. 24 of \cite{Ishkhanyan09}. 
%Even though on the scale displayed in Fig.\ \ref{fig4} the agreement 
%between Eq.\ \ref{multiple} and our numerical results appears to be good, 
The analytical predictions for small $g$ match the numerical data well for the higher excitation-free points, 
but less so for the lower resonant velocities. The ratios of the slopes of $v_\mathrm{res}$ as a function of $g$ between the analytical and the numerical results are 0.47, 0.83, 0.92 and 0.95 for the first four excitation-free points in increasing velocity order.

% Experimental considerations.
An experimental setup to test these results can be implemented with ultracold Bose gases trapped on an atom chip. 
A $1d$ quasi condensate can fill a potential tube
of several hundred micron length \cite{Kruger10} with a very small potential variation along the tube 
and correspondingly homogeneous density. The gas can be set in motion at a
controlled velocity by removing the residual longitudinal confinement and applying 
a short pulse of a magnetic gradient in the same direction. Velocities on the order of and
exceeding a typical sound velocity of $c\sim 1$ mm/s can be achieved straightforwardly. 
By applying currents to microwires on the chip, a magnetic defect can be produced and
controlled. Its geometric shape can be tailored with a resolution given by the atom-surface distance $z_0$. Here it is critical that 
$z_0$ is on the order of single microns in order to distinguish the individual excitationless resonances from 
the intermittent regimes of fast excitation growth. The excitation behaviour can be probed by varying the velocity $v$ of 
the gas's motion or by varying the final amplitude of the defect $V_0$ at fixed $v$. 
The difference in $V_0$ for the first two excitation-free points is expected to be 
$\Delta V_0 \approx \frac{3 h^2}{2m}\frac{1}{d^2}=h \times 6.9$~kHz$\times \mu$m$^2\frac{1}{d^2}$ for the example of $^{87}$Rb, 
so that $d\gtrsim z_0$ needs to be sufficiently small to maintain $\Delta V_0\approx \mu \approx h\times$1 kHz, where $\mu$ is the chemical potential of the repulsively interacting gas. Appropriate rise times of the defect are on the order of milliseconds ($t_\mathrm{barrier}\gtrsim 1$ms) for $z_0\gtrsim 1\mu$m.

%Conclusions General understanding * Qualitative/conceptual of why the LFP exist - Other potentials

In conclusion, we have studied the propagation of matter waves across defects of rectangular shape starting 
from a stationary flow solution in the defect-free case with velocity $v$ and ramping on the defect: for velocities smaller 
than a critical velocity $v_c$ there is no production of excitations (with $v_c$ very well approximated by the Landau critical velocity 
$v_L$ in our one-dimensional case). For a set of supercritical velocities $v>v_c$ the growth of excitations is fully suppressed, contrary to the generic expectation as a consequence of the Landau criterion. For these velocities, we find the production of excitations to be 
bounded in time 
%(as for its temporal rate), 
and to stop entirely when the defect is completely turned on. The flow is stable for times far in excess of any typical experimental timescale in the domain of ultracold atoms and elongated Bose condensates in particular. The obtained findings do not crucially depend 
on the ramping time of the defect. Such excitation-free supercritical velocities are present both for wells and barriers, 
and for repulsive and (small) attractive interactions. We observe that even though in the nonlinear case bound 
states and bifurcation effects are expected, our protocol allows us to access the excitation-free points 
in a clean way, not depending on the ramping time.

The obtained excitation-free supercritical velocities are the nonlinear counterpart of the velocities having unit transmission in the linear Schr\"odinger case (Ramsauer-Townsend resonances) and are due to the resonance 
between the length scale associated with the matter wave momentum ($\sim 2\pi/k$) and the length scale of the defect. The shift 
from the resonance is positive (negative) for repulsive (attractive) interaction in the case of barrier defects, and vice  
versa for well defects. Beyond this surprising proof of the presence of such excitation-free supercritical velocities for the paradigmatic case of rectangular defects, we expect 
that they exist in general whenever a defect can be characterised by a well defined length scale, e.g.\ for trapezoidal defects or two delta-peaked potentials. The steeper the defect is at its edges, the more robust will the inhibition of excitations be in the vicinity of a set of supercritical velocities.

Motivated by our work it will be interesting to study general criteria of existence and stability of supercritical solutions of the 
nonlinear Schr\"odinger equation in presence of defects with a well defined length scale. On the other hand, intriguing possibilities arise from utilising supercritical 
points in measurement devices based on superfluids and superconductors. Tuning a barrier to a supercritical flow resonance would facilitate precise determination of unknown external parameters affecting the flow velocity such as rotation 
and performing selective measurements at supercritical velocities.\\ 

{\em Acknowledgements:} Discussions with S. Fagnocchi, G. Dell'Antonio, P. Kevrekidis , G. Mussardo and A. Recati are gratefully acknowledged. 
A.T. acknowledges support from the Italian PRIN 
``Fenomeni quantistici collettivi: dai sistemi fortemente correlati ai simulatori quantistici'' 
(PRIN 2010\_2010LLKJBX). A.P.M. acknowledges support from CONACyT. Support from the EU-FET Proactive Action (Grant 601180 ``MatterWave'') and EPSRC (Grant EP/I017828/1) is also acknowledged.

\end{document}